\documentstyle[aps,prb,amsmath,epsfig]{revtex}

\begin{document}

\title{Dynamic correlations of the Coulomb Luttinger liquid}
\author{Yasha~Gindikin and V.~A.~Sablikov}
\address{Institute of Radio Engineering and Electronics,
Russian Academy of Sciences, Fryazino, Moscow District, 141120, Russia}
\date{\today}
\maketitle
\begin{abstract}
The dynamic density response function, form-factor, and spectral function of a Luttinger
liquid with Coulomb electron-electron interaction are studied with the emphasis on the
short-range electron correlations. The Coulomb interaction changes dramatically the
density response function as compared to the case of the short-ranged interaction. The
form of the density response function is smoothing with time, and the oscillatory
structure appears. However, the spectral functions remain qualitatively the same. The
dynamic form-factor contains the $\delta$-peak in the long-wave region, corresponding to
one-boson excitations. Besides, the multi-boson-excitations band exists in the wave-number
region near to $2k_F$. The dynamic form-factor diverges at the edges of this band, while
the dielectric function goes to zero there, which indicates the appearance of a soft mode.
We develop a method to analyze the asymptotics of the spectral functions near to the
edges of the multi-boson-excitations band.
\end{abstract}

\section{Introduction}
Electron-electron (e-e) interaction is known to produce the most
pronounced effects in one-dimensional (1D) systems, where a
strongly correlated state appears even if the interaction is
weak.~\cite{LL} A number of attempts were made to treat electron
correlations in 1D using the approaches, devised for 2D and 3D.
They are based on the random phase approximation (RPA) with various
versions of the local field corrections, both static and
dynamic.~\cite{Mahan,STLS,QSTLS} These methods give no
qualitatively new results as compared to higher dimensions, with
the picture of charge excitations being exhausted by common
long-wave plasmons.~\cite{Gold,DasSarmaIII,Agosti} However, this
way to consider electron correlations has obvious shortcomings,
especially in what concerns the dynamic short-range correlations,
inherent in 1D. The flaws of the RPA-like approaches become worse
as the dimensionality is reduced.~\cite{Johnson} In 1D there appear
such unphysical results as the negative dynamic
form-factor,~\cite{Tanatar} the violation of the compressibility
sum rule and negative pair correlation function, the latter two
being also present in 2D and 3D.~\cite{STLS,QSTLS}

Presently, the most advanced way to treat dynamic electron correlations in 1D is based on the
Luttinger liquid (LL) theory.~\cite{LL,TL,ML,LP,HaldaneII} The short-range
correlations~\cite{Emery,Haldane,SG} appear here in a natural way, because this model
takes properly into account the nesting of the Fermi surface for the wave number equal to
$2k_F$, with $k_F$ being the Fermi wave number. Owing to the short-range correlations, the LL
density response contains the $2k_F$ component, in addition to the usual long-wave one. In
what follows the $2k_F$ term is also referred to as the charge-density wave (CDW).

The $2k_F$ susceptibility $\chi_{\rm CDW}(q,\omega)$ was considered by Luther and
Peschel~\cite{LP} for the LL with short-ranged e-e interaction. They have found that
$\chi_{\rm CDW}(q,\omega)$ diverges as $|q-2k_F|\to \omega/v$, where $v$ is the velocity
of the LL bosonic excitations. We stress that the divergency of the susceptibility is
very important because it corresponds to the presence of a collective mode in the system.

In real 1D structures e-e interaction is, generally speaking, long-ranged, unless it is
screened by a metallic gate. It is well known that Coulomb interaction can highly
modify the ground state and transport properties of 1D systems. Thus, in a Luttinger
liquid with Coulomb interaction (CLL) the static correlations become much stronger than in
the short-ranged LL, with the short-range $2k_F$ ($4k_F$ in a spinful case) component
being the dominating one.~\cite{Schulz}

How the dynamic LL correlations are affected by Coulomb interaction is the highly
intriguing question. However, considering only the long-wave density component
does not lead to the qualitatively new behaviour of the dynamic form-factor and collective
modes.~\cite{LL,VoitII,Kramer,SS,DasSarmaII} The results are very close to those obtained
in RPA.

The present paper concentrates on the CDW contribution to the
dynamic response functions and spectral characteristics of the
Coulomb Luttinger liquid. We show analytically that Coulomb
interaction strengthens the divergency of the dynamic density
susceptibility in comparison with that of the short-ranged LL. The
singularity of the CDW susceptibility leads to the new behaviour of
the collective charge mode that is immanent only in 1D. The strong
spatial dispersion arises in the vicinity of $q=2k_F$, the mode
frequency going to zero as $q\to2k_F$. This means that the soft
mode appears. Such soft mode is absent in 2D and 3D systems,
because the short-range dynamic electron correlations are much
weaker there. Note that the $2k_F$ mode can not be obtained within
the RPA approach, even with local field corrections.

Our results concerning the CLL spectral function diverge from those of
Ref.~\onlinecite{DasSarma}, where it was argued that the non-linear dispersion of bosons,
which appears in the presence of Coulomb interaction, kills the spectral function
singularities. Instead, a flattened maximum was found in a spectral function, the position
of the maximum being shifted from the resonant frequency. We show that the Coulomb
interaction, on the contrary, only strengthens the divergency, which is a rather general
result for the CLL correlation functions.

The outline of the paper is as follows. In Sec.~II we investigate
the dynamic density response function, form-factor, spectral
function, dissipative conductance, and dielectric function of a
CLL, comparing them with the short-ranged LL results. In Appendices~A
and~B the methods to calculate the dynamic form-factor of a
short-ranged LL and CLL are presented.

\section{Dynamic correlations}
We start with a bosonized spinless LL Hamiltonian\cite{LL}
$$H=\sum_p\hbar\omega_pb^+_pb_p,$$
where $b^+_p(b_p)$ are boson creation (annihilation) operators. The boson frequency is
given by $\omega_p=|p|v(p)$, where the velocity of excitations is $v(p)=v_F/g(p)$, with
the interaction parameter $g(p)$, the Fermi velocity $v_F$. For the short-range
interaction $g$ is constant ($0<g<1$ for e-e repulsion), whereas for Coulomb interaction
in the long-wave limit $|pd|\ll 1$ the interaction parameter is $g(p)=\beta
|\ln|pd||^{-1/2}$, $d$ being the quantum wire diameter, $\beta=[\pi
\hbar v_F/2e^2]^{1/2}$.

The electron-density-fluctuation operator in the Luttinger model is written
as~\cite{Haldane,SG}
\begin{equation}
\rho(x)=-\dfrac{1}{\pi}\partial_x\phi+\dfrac{1}{2\pi}\partial_x\sin(2k_Fx-2\phi),
\label{density}
\end{equation}
$\phi(x)$ being the bosonic phase. The first component of the density operator, $\rho_{\rm
lw}$, describes long-wave excitations and represents the sum of the densities of the
right- and left-moving electrons. The corresponding excitations come about separately
within each branch of electron spectrum and have momentum $q\ll k_F$. The second
component, $\rho_{\rm CDW}$, which rapidly oscillates in space, is due to the interference
of the right- and left-moving electrons. This corresponds to excitations with momentum
$q\approx 2k_F$. It is this term that describes the short-range electron correlations.
Note that the presented form of $\rho_{\rm CDW}$ differs from the conventional
one~\cite{LL} in that the former is the exact differential. The exact differential form of
the density fluctuation operator guarantees the particle number conservation in an
isolated 1D system. The conventional form of $\rho_{\rm CDW}$ does not conserve the number
of particles and thus violates the electroneutrality of the 1D system.~\cite{SG}

\subsection{The density response function}

\begin{figure}
\begin{center}
\epsfig{file=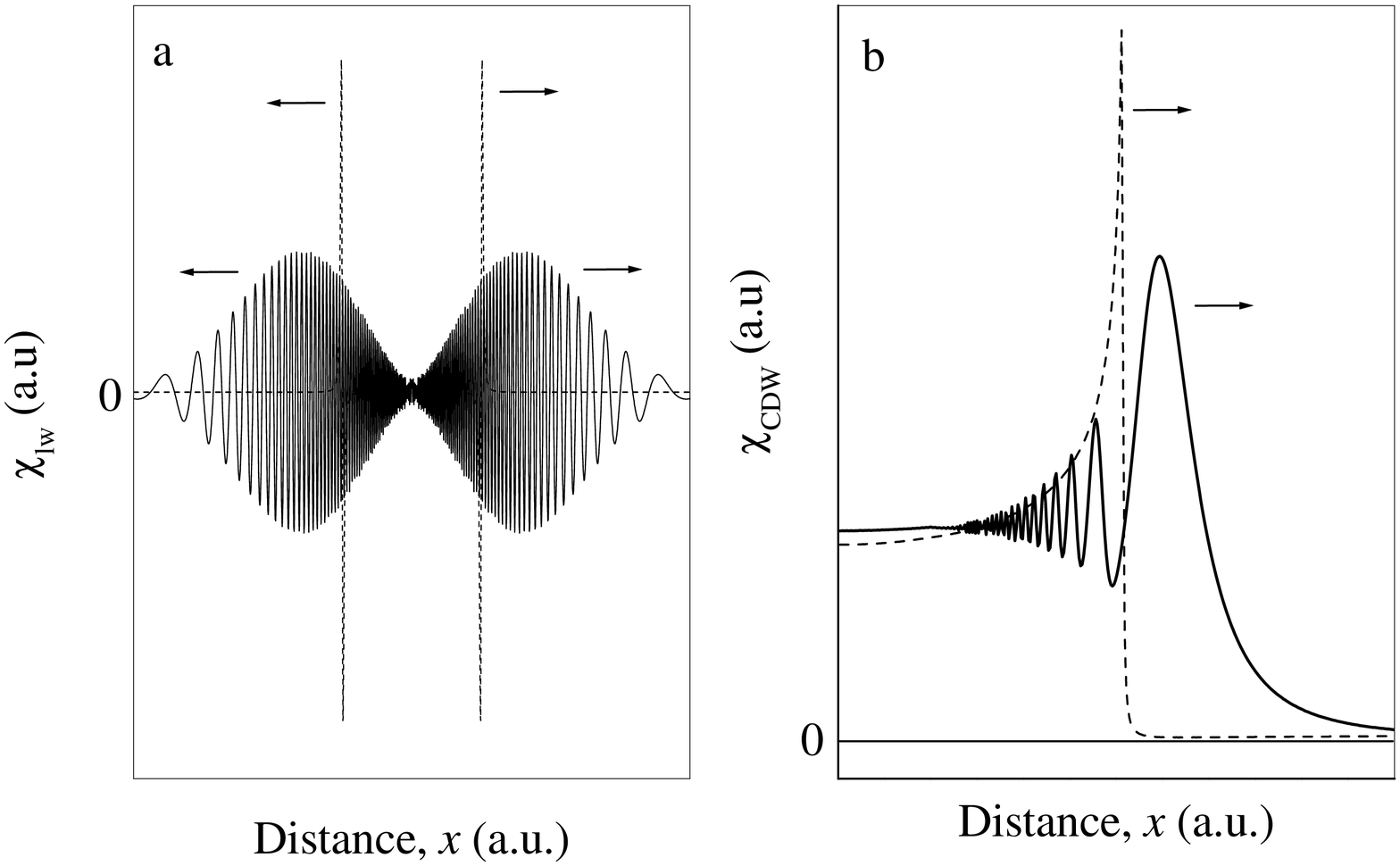, width=14cm}
\caption[to]{The long-wave (a) and  CDW (b) components of the density
response function for the short-ranged LL (dash line) and CLL (solid line). In the CDW
response function the $2k_F$ filling is not shown. Arrows show the direction of wave
propagation.}
\label{DRF}
\end{center}
\end{figure}

For the two density components of Eq.~(\ref{density}), the density response function (DRF)
is calculated via the Kubo formula to give:
$$\chi(x,t)=\chi_{\rm lw}(x,t)+\chi_{\rm CDW}(x,t),$$ where
$$
\chi_{\rm lw}(x,t)=\dfrac{\theta(t)}{\pi h}\partial^2_{x}f_2(x,t),
$$
$$
\chi_{\rm CDW}(x,t)=\dfrac{\theta(t)}{2\pi h}\partial^2_{x}\left[e^{-
f_1(x,t)}\sin(f_2(x,t))\cos(2k_Fx)\right].
$$
The functions $f_1(x,t)$ and $f_2(x,t)$ are as follows:
\begin{equation}
f_1(x,t)=v_F\int_0^{+\infty}\dfrac{dp}{\omega_p}\left[2-\cos(\omega_pt+px)-\cos(\omega_pt-
px)\right]e^{-\alpha p}
\label{f1}
\end{equation}
\begin{equation}
f_2(x,t)=v_F\int_0^{+\infty}\dfrac{dp}{\omega_p}\left[\sin(\omega_pt+px)+\sin(\omega_pt-
px)\right]e^{-\alpha p},\;
\alpha=k_F^{-1}.
\label{f2}
\end{equation}
For the short-ranged e-e interaction these functions can be calculated
exactly:~\cite{LP,SG,VoitII}
$$e^{-f_1(x,t)}=\alpha^{2g}[v^2t^2-x^2]^{-g}, $$
$$f_2(x,t)=\pi g \theta(v^2t^2-x^2).$$
The DRF behaviour is illustrated by Fig.~\ref{DRF} for both short-ranged and Coulomb LLs.

As is seen, the DRF of a short-ranged LL is presented by two wave fronts, propagating in
opposite directions with constant velocity $v$. The Coulomb DRF, as distinct from the
short-ranged one, has no sharp fronts. The wave form is smoothing because of the nonlinear
dispersion of CLL boson excitations. The characteristic space scale of the Coulomb DRF
depends on time as $x\approx t\sqrt{\ln t}$, where dimensionless $x$ and $t$ are
normalized, respectively, by $d$ and $\beta d/v_F$. The non-periodic oscillatory structure
of the DRF arises in the CLL because the phase velocity diverges at wave number $q=0$.
This divergency leads to the appearance of the stationary phase $\phi_{\rm st}$ in
integrals~(\ref{f1}),~(\ref{f2}). The oscillatory structure, owing to the strong Coulomb
dispersion, complements the smoothing of the wave form, usual for the wave propagation in
weakly dispersive media. The asymptotic decay of the Coulomb DRF with the time is
extremely slow, and has the form similar to the decay of the static CLL correlator with
the distance.~\cite{Schulz} Devoid of a sharp front, the Coulomb DRF contains a slowly
decaying tail at large distance $x\gg t\sqrt{\ln t}$.

Introducing the characteristic distance $x_f=t\sqrt{\ln t}$ and stationary phase
$$\phi_{\rm st}=-\dfrac{t}{2\sqrt{e}}\left(\sqrt{2}-
2\dfrac{x}{t}+\dfrac{x^2}{\sqrt{2}t^2}+O\left(\dfrac{x^3}{t^3}\right)\right),$$ the
asymptotic behavior of $f_1$ and $f_2$ can be presented as follows:

i) $x\gg x_f$
$$f_1(x,t)\sim 2\beta\left(\sqrt{\ln(x-x_f)}+\sqrt{\ln(x+x_f)}\right),$$
$$f_2(x,t)\sim \dfrac{\pi \beta}{2}\frac{x_f^3}{x^3},$$

ii) $x\ll x_f$
$$f_1(x,t)\sim 2\beta\left[\sqrt{\ln(x_f-x)}+\sqrt{\ln(x_f+x)} -\sqrt{\dfrac{\pi
\sqrt{2e}}{2t}}\cos(\phi_{\rm st}+\frac{\pi}{4})\right],$$
$$ f_2(x,t)\sim 2\beta\left[\dfrac{1}{\sqrt{\ln(x_f-x)}}+\dfrac{1}{\sqrt{\ln(x_f+x)}}-
\sqrt{\dfrac{\pi
\sqrt{2e}}{2t}}\sin(\phi_{\rm st}+\frac{\pi}{4})\right].$$

We draw attention to the fact that in 1D there are two different
mechanisms of the density evolution.~\cite{SG} First of all, there
exist sound-like waves, caused by forward scattering and described
by the long-wave DRF component. In this kind of motion neighboring
electrons move almost in the same phase, so that the corresponding
correlations are almost static. Secondly, electrons suffer the
backward scattering from the nearest particles and interfere, which
gives rise to $2k_F$ density oscillations. Electron correlations
that are related to the $2k_F$-mechanism of electron density
response are essentially dynamic. Therefore, taking into account
the short-range electron correlations, the two-particle Wigner function
$f(x_1,p_1,x_2,p_2,t)$ in no way can be represented as a product of
two single-particle Wigner functions and the {\it static} pair
correlation function $g(x)$, that is
$$ f(x_1,p_1,x_2,p_2,t)\ne f(x_1,p_1,t) f(x_2,p_2,t)\,g(x_1-x_2)$$
in 1D. Thus the basic assumption of the
Singwi-Tosi-Land-Sj\"{o}lander theory~\cite{STLS,QSTLS} is violated
in 1D, which explains why the $2k_F$ collective mode is overlooked
in the excitation spectrum within this approach.

\subsection{The form-factor}
Now we turn to the LL dynamic form-factor $S(q,\omega)$, which is the Fourier transform of
the density-density correlator $R(x,t)=\langle\rho(x,t)\rho(0,0)\rangle$. At zero
temperature the form-factor coincides with the imaginary part of the susceptibility:
$S(q,\omega)=-2\hbar \chi''(q,\omega).$ Although the form of the Coulomb DRF $\chi_{\rm
lw}(x,t)$ differs dramatically from the short-ranged one, the long-wave part of the
form-factor, $S_{\rm lw}(q,\omega)$, has the simple universal expression~\cite{LL} for any
dispersion law $\omega_q$:
$$ S_{\rm lw}(q,\omega)=|q|g(q)\delta(\omega-\omega_q).$$
Thus, only the position of the peak is shifted in a CLL. This is clear physically, since
$S_{\rm lw}(q,\omega)$ determines the probability to create a {\it single} boson when
absorbing the quantum $\hbar \omega$, which fixes the singularity position.

The CDW form-factor $S_{\rm CDW}(q,\omega)$ describes the excitation of several bosons and
is much more interesting. For the short-range interaction $S_{\rm CDW}(q,\omega)$ can be
exactly calculated:\cite{LP,SG,VoitII}
\begin{equation}
S_{\rm
CDW}(q,\omega)=\dfrac{1}{v_F}\dfrac{g}{4^{g+1/2}\Gamma^2(g)}\left(\dfrac{q}{k_F}\right)^
2\sum_{r=\pm 1}
\left[\left(g\dfrac{\hbar\omega}{\varepsilon_F}\right)^2-\left(\dfrac{q}{k_F}-
2r\right)^2\right]^{g-1},
\label{FSR}
\end{equation}
$\varepsilon_F$ being the Fermi energy. The entire complex susceptibility is
\begin{equation}
\chi_{\rm CDW}(q,\omega)=-\dfrac{1}{\hbar v_F}\dfrac{g}{4^{g+1}\Gamma^2(g)\sin(\pi
g)}\left(\dfrac{q}{k_F}\right)^2\sum_{r=\pm 1}
\left[\left(\dfrac{q}{k_F}-2r\right)^2-
\left(g\dfrac{\hbar\omega+i0}{\varepsilon_F}\right)^2\right]^{g-1}.
\label{HSR}
\end{equation}
The details of the calculation are presented in Appendix A. Notice that the CDW
form-factor is zero out of the $q$ band $||q|-2k_F|<g\omega/v_F$ and diverges at the band
edges. For simplicity, in what follows we consider $q>0$.

For the CLL the CDW form factor can not be calculated exactly, but its general properties
are easily understood, using the formula
$$
S_{\rm CDW}(q,\omega)=(2\pi)^2\sum_m|\langle m|\rho_{\rm
CDW}|0\rangle|^2\delta(\omega-\omega_m)\delta(q-q_m-2k_F),
$$
which can be obtained directly from the expression~(\ref{R}) for the CDW density
correlator. This formula differs from the conventional one~\cite{Zhen'ka} in that the wave
number argument of the $\delta$-function is additionally shifted by $2k_F$. The
conventional derivation of the form-factor representation via $\delta$-functions pays no
attention to the $2k_F$ modulation of the density operator of Eq.~(\ref{density}).

The sum in the last expression is taken over all stationary $|m\rangle$ states of the
system, $\hbar \omega_m$ being the state energy and $\hbar q_m$ - the state momentum. The
state consists of a number of bosons, excited above the vacuum. The specific form of the
matrix element is not of interest now, but what is important is that $\langle m|\rho_{\rm
CDW}|0\rangle$, in contrast to $\langle m|\rho_{\rm lw}|0\rangle$, is non-zero for excited
states, containing more than one boson. Thus all possible boson systems of the total
energy $\hbar \omega$ and momentum $\hbar (q-2k_F)$ contribute to $S_{\rm CDW}(q,\omega)$.

The boson dispersion curve $\omega=\omega_p$ is convex, i.e. $\omega ''_p<0$. Therefore,
if the boson system has the total momentum $\hbar p$, then its energy can not be less than
$\hbar
\omega_p$. Whence, the form-factor $S_{\rm CDW}(q,\omega)$ is zero
when $\omega<\omega_{q-2k_F}$. When $\omega=\omega_{q-2k_F}$, only one boson can be
excited. As one increases $\omega$ from the threshold value $\omega_{q-2k_F}$, the number
of different boson systems of the given total energy $\hbar
\omega$ and momentum $\hbar (q-2k_F)$ increases rapidly, and their
contribution to the form-factor is increasing too. The formation of boson systems occurs
when $\omega$ is being shifted from the threshold on the scale $v_F/L$, where $L$ is the
length of the system. Hence, taking the thermodynamic limit, i.e. $L\to \infty$, we find the
non-zero values of $S_{\rm CDW}(q,\omega)$ as $\omega \to
\omega_{q-2k_F} +0$. Moreover, we show in Appendix B that
$S_{\rm CDW}(q,\omega)$ diverges as $\epsilon=\omega-
\omega_{q-2k_F} \to +0$, just like in the short-ranged LL:
\begin{equation}
\label{FF}
\omega\, S(q, \omega)\sim \dfrac{e^{-4\beta|\ln \epsilon|^{1/2}}}{\epsilon |\ln
\epsilon|^{1/2}}.
\end{equation}

\begin{figure}
\begin{center}
\epsfig{file=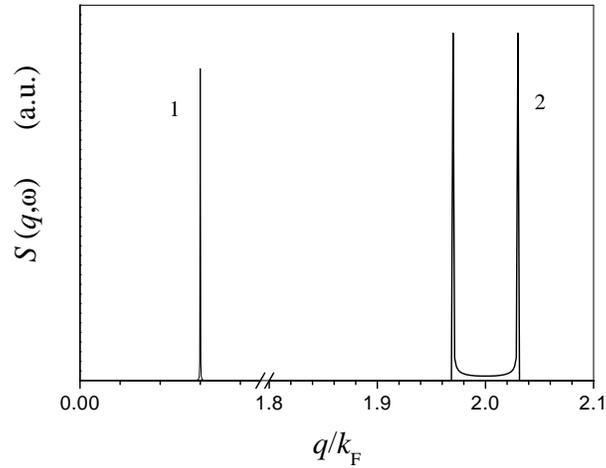, width=10cm}
\caption[to]{The CLL dynamic form-factor as a function of wave number $q$.
The $\delta$-peak in the long-wave region corresponds to one-boson
excitations. The multi-boson-excitations band exists in the
vicinity of $q=2k_F$.}
\label{formf}
\end{center}
\end{figure}

The CLL dynamic form-factor, containing both CDW and long-wave components, is shown in
Fig.~\ref{formf} as a function of $q$ at fixed $\omega$.

In a similar way we find that the CLL spectral function~\cite{LL} is zero when
$\omega<\omega_q$ and diverges as $\delta=\omega -\omega_q \to +0$:
\begin{equation}
\omega \,\rho(q, \omega)\sim \dfrac{e^{-A\beta|\ln \delta|^{1/2}}}{\delta |\ln
\delta|^{1/2}},
\label{spect}
\end{equation}
with $A$ being $(\dfrac{1}{g(q)}-1)^2$.

The last result contradicts to the one obtained in
Ref.~\onlinecite{DasSarma}, where it was claimed that the nonlinear
boson dispersion flattens the singularity of $\rho(q,\omega)$ at
$\omega=\omega_q+0$, producing a maximum instead, with
$\rho(q,\omega)$ going to zero at the resonant frequency. The
approach of Ref.~\onlinecite{DasSarma}, if applied to the CDW
form-factor, would force us to conclude that the CLL form-factor
has a cusp instead of a singularity. In the following subsection we
argue that such conclusion is physically incorrect.

Notice that since the expression~(\ref{FF}) represents the exact differential, the
form-factor singularity is integrable, as it should be, because the integral of $S_{\rm
CDW}(q,\omega)$ with respect to $\omega$ gives the static form-factor $S(q)$, which is
finite at non-zero wave numbers. The static form-factor was calculated via direct
integration of the static density-density correlator to give
$$
S(q)\sim \dfrac{e^{-4\beta|\ln q|^{1/2}}}{q |\ln q|^{1/2}}.
$$

The dynamic form-factor determines the power $P$ that is dissipated in a LL, disturbed by
external electric potential~$\varphi$:
$$P(\omega)=\dfrac{e^2}{4h}\int_{-\infty}^{+\infty}\omega S(q, \omega)|\varphi(q)|^2dq.$$

The contribution of the long-wave density response to the dissipated power was
investigated in detail in Ref.~\onlinecite{Kramer}. It was shown there that the dissipated
power determines the conductance of a LL, providing that no current-carrying leads are
taken into account. The CDW contribution to the dissipated power was calculated in
Ref.~\onlinecite{SG} for the short-ranged LL. In a CLL, the frequency dependence of
$P_{\rm CDW}(\omega)$ is as follows:
$$P_{\rm CDW}(\omega)\sim\dfrac{e^{-4\beta |\ln \omega|^{1/2}}}{|\ln
\omega|^{1/2}}|\varphi(2k_F)|^2,$$
where $\omega$ is normalized by $v_F/\beta d$, so that the dimensionless $\omega \ll1$.
For the short-ranged LL the frequency dependence is:
$$P_{\rm CDW}(\omega)\sim \omega^{2g}|\varphi(2k_F)|^2.$$
In a CLL the power $P_{\rm CDW}$ diminishes very slowly with $\omega$, so that in the
low-frequency regime the CDW dominates in the dissipation,\cite{SG} since the power
dissipated due to the long-wave density component behaves as $P_{\rm lw}\sim \omega^2$.

\subsection{Dielectric function}
\begin{figure}
\begin{center}
\epsfig{file=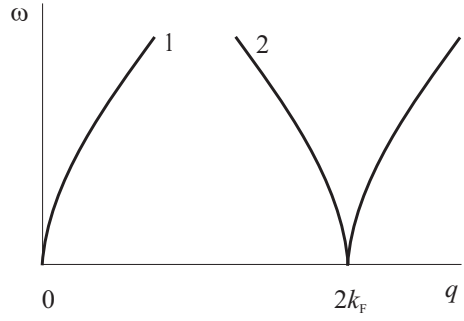, width=8cm}
\caption[to]{The dispersion of collective modes in a CLL. Line (1) is a plasmon mode. Line (2)
is $2k_F$ mode, related to the short-range electron correlations. }
\label{modes}
\end{center}
\end{figure}

The dielectric function $\varepsilon(q,\omega)$ is connected with the dynamic
susceptibility via
$$\varepsilon^{-1}(q,\omega)=1+V(q)\chi(q,\omega),$$
$V(q)$ being the e-e interaction potential. The dielectric function evidently goes to zero
when the form-factor goes to infinity. The collective modes of the system under
considerations are determined by zeros of $\varepsilon(q,\omega)$. Since the form-factor
has singularities in two wavenumber regions, we find two regions, where the collective
modes can propagate. The dispersion of the collective modes in a LL is illustrated by
Fig.(\ref{modes}).

The first region corresponds to the singularity of the long-wave form-factor component
$S_{\rm lw}(q,\omega)$, which occurs at $\omega=\omega_q$. This mode is just long-wave
plasmons, almost identical to the RPA ones.~\cite{Mahan,Kramer,DasSarmaII}

The singularities of the CDW form-factor $S_{\rm CDW}(q,\omega)$ give the new mode,
situated near to $q=2k_F$. The dispersion of this mode is as follows: $\omega=\omega_{q-
2k_F}$. Here $\omega$ is pure real, which means that the $2k_F$ mode is non-decaying. This
is because the form-factor $S_{\rm CDW}(q,\omega)$ has the true divergency. The cusp
instead of a singularity in $S_{\rm CDW}(q,\omega)$ would lead to the complex frequency
solutions of the equation $\varepsilon(q,\omega)=0$ and thus to the strong damping of the
collective mode. We believe that such damping is physically absurd. Indeed, the boson
excitations are non-interacting in the frame of the Luttinger model. They can not decay
into Landau quasi-particles either. Hence, there is no possibility for a collective mode
in a CLL to transfer its energy to some other excitations and thus to damp. This is clear
with the bosonization approach, which explains also why the usual long-wave plasmons are
not damping in 1D.

An important conclusion is that the mode frequency goes to zero at $q\to 2k_F$, in other
words, the mode is soft. It is the presence of the $2k_F$ mode that principally
distinguishes the LL picture of collective excitations from the RPA one.

\section{Conclusion}
In the present work we have investigated the response functions of a spinless Luttinger
liquid with Coulomb interaction at zero temperature. We have found that:

i) The CLL density response function is qualitatively different from that of a
short-ranged LL. The nonlinear dispersion of bosonic excitations in a CLL results in that
the DRF has no sharp front, and as the wave propagates, its form is smoothing with the
appearance of the oscillatory structure.

ii) The dynamic CDW form-factor is non-zero only in a region where
$\omega_{q-2k_F}<\omega$, which is a consequence of energy and momentum conservation laws,
applied to LL bosons. The CDW form-factor is diverging as $\epsilon=\omega-
\omega_{q-2k_F}\to +0$ like $S(q, \omega)\sim \exp({-4\beta|\ln \epsilon|^{1/2}})/\epsilon |\ln
\epsilon|^{1/2}$. The similar result was obtained for the spectral
function.

iii)  Owing to the CDW contribution, the dielectric function goes to zero as $\omega
\to \omega_{q-2k_F}+0$. This means that the non-decaying mode appears in the
region near to $q=2k_F$. The frequency of the mode tends to zero as $q\to 2k_F$, which
means that the mode is soft. This $2k_F$ mode exists because the short-range dynamic
electron correlations are highly pronounced in 1D, which is due to the $2k_F$ nesting of
the Fermi surface of 1D electrons. RPA-like approaches are unable to describe the $2k_F$
mode since the dynamic nature of the short-range correlations is neglected there.

\section*{Acknowledgments}
It is a pleasure to thank V.~A.~Volkov for helpful discussions.
This work was supported by the Russian Fund for Basic Research (Grant No 99-02-18192),
Russian Program "Physics of Solid-State Nanostructures" (Grant No 97-1054) and Russian
Program "Surface Atomic Structures" (Grant No 5.3.99).

\appendix
\section{The CDW form-factor of the short-ranged LL}
In this section we calculate the CDW form-factor and the entire complex CDW susceptibility
of a short-ranged LL. Consider the CDW density correlator:
\begin{equation}
R_{\rm CDW}(x,t)=-\dfrac{1}{8\pi ^2}\partial_x^2\left(\exp\left[-v_F\int_{-
\infty}^{+\infty}\dfrac{dp}{\omega_p}\left(1-e^{-i\omega_pt-ipx}\right)e^{-
\alpha\left|p\right|}\right]\cos\left(2k_Fx\right)\right).
\label{R}
\end{equation}
The form-factor is, by definition, the Fourier tranform of $R(x,t)$:
\begin{equation}
S(q,\omega)=\int_{-\infty}^{+\infty}dx\,e^{iqx}\int_{-\infty}^{+\infty}dt\,e^{i\omega
t}R(x,t).
\label{d}
\end{equation}
First of all, we reduce the calculation of the $S_{\rm CDW}(q,\omega)$ to the calculation
of a simpler function $F(q,\omega)$. Substituting Eq.~(\ref{R}) into Eq.~(\ref{d}), we get
\begin{equation}
S_{\rm CDW}(q,\omega)=\dfrac{q^2}{16\pi^2}\left(F(q-2k_F,\omega)+ F(q+2k_F,\omega)\right),
\label{Saux}
\end{equation}
where
\begin{equation}
F(q,\omega)= \int_{-\infty}^{+\infty}dx\,e^{iqx}\int_{-\infty}^{+\infty}dt\,e^{i\omega
t}F(x,t),
\label{bas}
\end{equation}
and
\begin{equation}
F(x,t)= \exp\left(-v_F\int_{-\infty}^{+\infty}\dfrac{dp}{\omega_p}\left(1-e^{-i\omega_pt-
ipx}\right)e^{-\alpha \left|p\right|}\right).
\label{sab}
\end{equation}
For the short-range interaction the dispersion is linear: $\omega_p=v|p|$, and the
function $F(x,t)$ is easily calculated to give
$$
F(x,t)=\alpha^{2g}\left[\alpha+i(vt-x)\right]^{-g}\left[\alpha+i(vt+x)\right]^{-g}.
$$
Denote $\xi=vt-x$, $\zeta=vt+x$ to get
$$F(q,\omega)=\dfrac{\alpha^{2g}}{2v}\int_{-\infty}^{+\infty}d\xi e^{ia\xi}(\alpha+i\xi)^{-
g}\int_{-\infty}^{+\infty}d\zeta e^{ib\zeta}(\alpha+i\zeta)^{-g},$$ where
$a=\dfrac12\left(\dfrac{\omega}{v}-q\right)$, and
$b=\dfrac12\left(\dfrac{\omega}{v}+q\right)$. Thus we have reduced the double
Fourier-tranform to the single one. The integrals are easily calculated by closing
contours upwards or downwards, depending on $a$ and $b$ signs, to give
$$F(q,\omega)=\dfrac{8\pi^2}{v}\left(\dfrac{\alpha^{g}}{2^g\Gamma(g)}\right)^2\left(\dfrac{
\omega^2}{v^2}-q^2\right)^{g-1}\Theta\left(\dfrac{\omega^2}{v^2}-q^2\right).$$
Substituting the last expression into Eq.~(\ref{Saux}) finally leads to Eq.~(\ref{FSR}).
Now we have the imaginary part of the susceptibility. How can we restore the entire
complex $\chi(q,\omega)$, avoiding the direct use of Kramers-Kronig relations? We propose
to guess it:
$$\chi_{\rm CDW}(q,\omega)=-\dfrac{1}{\hbar v_F}\dfrac{g}{4^{g+1}\Gamma^2(g)\sin(\pi
g)}\left(\dfrac{q}{k_F}\right)^2\sum_{r=\pm 1}
\left[\left(\dfrac{q}{k_F}-2r\right)^2-
\left(g\dfrac{\hbar\omega+i0}{\varepsilon_F}\right)^2\right]^{g-1}+({\rm unknown}\;\chi_1).$$
Indeed, the first term on the R.H.S. correctly gives the imaginary part of
$\chi(q,\omega)$. Thus the unknown function $\chi_1$ is a pure real function. On the other
hand, $\chi_1$ is analytic in the upper half-plane of $\omega$ (because both the L.H.S.
and the first term on the R.H.S. are analytic functions in the upper half-plane of
$\omega$, and $\chi_1$ is their difference). Whence, $\chi_1=0$ as a consequence of
Kramers-Kronig relations. This way we get Eq.~(\ref{HSR}).

\section{The CDW form-factor of the Coulomb LL}
Here we present a method to find $S_{\rm CDW}(q,\omega)$ for the CLL in the region near
the threshold $\omega=\omega_{q-2k_F}$. In the Coulomb case it is not possible to
calculate exactly the double Fourier transform of CDW density correlator, given by
Eq.~(\ref{R}). Instead, we propose to formulate the equation on $S(q,\omega)$, which would
contain only the spectral parameters $q$ and $\omega$. Then we solve this equation near
the threshold $\omega=\omega_{q-2k_F}$.

First of all, just like in Appendix A, we reduce the calculation of the $S_{\rm CDW}(q,\omega)$ to the calculation
of a simpler function $F(q,\omega)$. Substituting Eq.~(\ref{sab}) into Eq.~(\ref{bas}) and
performing once integration by parts with respect to $t$, we get our main equation:
\begin{equation}
\dfrac{\omega}{v_F}F(q,\omega)=\int_{-\infty}^{+\infty}dQ\,F(q-Q,\omega-\omega_Q).
\label{main}
\end{equation}
We stress that this equation contains only spectral-parameter dependence. The advantage of
this equation is that the resonant frequencies turn out to be specified here, and it is
much easier to extract the information about the spectral dependence of the form-factor
from this equation, rather than from the direct expression of Eq.~(\ref{bas}).

Let us shift the integration variable $Q$ by $q$, so that
\begin{equation}
\dfrac{\omega}{v_F}F(q,\omega)= \int_{-\infty}^{+\infty}dQ\,F(Q,\omega-\omega_{q+Q}).
\label{preexp}
\end{equation}
Here we used the fact that $F(q,\omega)$ is an even function of $q$ (since $R(x,t)$ is an
even function of $x$). Now let us expand the R.H.S. of Eq.(\ref{preexp}) w.r.t. $Q$ that
is contained in the frequency argument $(\omega-\omega_{q+Q})$:
\begin{equation}
\dfrac{\omega}{v_F}F(q,\omega)= \int_{-\infty}^{+\infty}dQ\,\left[F(Q,\omega-
\omega_{q})+\dfrac{Q^2}{2!}F_{qq}(Q,\omega-\omega_q)+ . . .\right].
\label{expansion}
\end{equation}
This expansion gives us all the necessary information.

i) Since the function $F(x,t)$ is analytic in the lower half-plane of complex time $t$
(which is seen from Eq.~(\ref{sab})), we find that $F(q,\omega)=0$ when $\omega<0$. (It
is, of course, clear physically, why the form-factor at zero temperature is zero when
$\omega<0$: there is no possibility to create an excitation below the ground state.) So,
the R.H.S. of Eq.~(\ref{expansion}) is zero when $\epsilon=\omega-\omega_q<0$.

Whence, the L.H.S., i.e. $F(q,\omega)$ has the threshold: when $\epsilon>0$, $F(q,\omega)$
is non-zero, whereas for $\epsilon<0$, $F(q,\omega)=0$. The existence of this threshold
was also explained from the physical background in the main text.

ii) The first term on the R.H.S., i.e. the integral
$\int_{-\infty}^{+\infty}dQ\,F(Q,\epsilon)$ equals $2\pi F_{\rm static}(\epsilon)$, where
\begin{equation}
F_{\rm static}(\epsilon)=\int_{-\infty}^{+\infty}dt\,e^{i\epsilon t}F(x=0,t).
\label{dst}
\end{equation}

Using integration by parts, all the terms on the R.H.S. of Eq.~(\ref{expansion}) can be
expressed via the Fourier-transform of the functions, depending on $t$ only. So, the
expansion (\ref{expansion}) allows one to reduce the calculation of the dynamic,
two-argument-dependent function $F(q,\omega)$ to the calculation of static functions,
depending on $\epsilon$ only. The calculation of these static functions takes {\it one}
integration only, and is much more easily performed. Using the direct calculation, it can
be shown that the function $F_{\rm static}(\epsilon)$ diverges as $\epsilon\to 0$, and so
do all the terms on the R.H.S. of Eq.~(\ref{expansion}). For a short-ranged LL, all the
terms diverge as a power-law of $\epsilon$ with the same exponent, therefore it would take
us to sum all the expansion to get the correct exponent on the L.H.S. In a CLL, the first
term divergency $F_{\rm static}(\epsilon)$ is the strongest one, and the following terms
divergencies were found to be weaker. So, in the leading order of divergency,
$F(q,\omega)$ in the CLL is given by
\begin{equation}
\dfrac{\omega}{v_F}F(q,\omega)\sim 2\pi F_{\rm static}(\omega-\omega_q),\;\;\omega-
\omega_q\to0.
\label{red}
\end{equation}
Using the asymptotic form of $F(x=0,t)$ at large $t$, we can find (via the direct
integration) the form of $F_{\rm static}(\epsilon)$ as $\epsilon\to0$. But we prefer
to get the result by another way. It is easy to show that $F_{\rm static}(\epsilon)$
satisfies the integral equation, similar to Eq.~(\ref{main}):
\begin{equation}
\dfrac{\omega}{v_F} F_{\rm static}(\omega)=2\int_0^{+\infty}dQ\, F_{\rm static}(\omega-
\omega_Q).
\end{equation}
Denote $\xi=\omega-\omega_Q$ to rewrite the last equation in the form
\begin{equation}
\dfrac{\omega}{v_F} F_{\rm static}(\omega)=2\int_0^{\omega}\dfrac{\beta
d\xi}{v_F\sqrt{\ln\dfrac{\omega-\xi}{v_F}\beta d}} F_{\rm static}(\xi).
\label{stat}
\end{equation}
The limits of integration are dictated by the property that $ F_{\rm static}(\omega)=0$ at
$\omega<0$. It is a good approximation to replace the kernel $\sqrt{\ln\dfrac{\omega-
\xi}{v_F}\beta d}$ with $\sqrt{\ln\dfrac{\omega}{v_F}\beta d}$, because the singularity of
$F_{\rm static}(\xi)$ at $\xi=0$, which gives the major contribution to the R.H.S. of
Eq.~(\ref{stat}), is then integrated with a correct weight. Then, multiplying both sides
of Eq.~(\ref{stat}) on $\sqrt{\ln\dfrac{\omega}{v_F}\beta d}$ and then differentiating
with respect to $\omega$, we get the following differential equation for $F_{\rm
static}(\omega)$:
$$
\dfrac{d}{d\omega}\left(\omega \sqrt{\ln\dfrac{\omega}{v_F}\beta d}F_{\rm
static}(\omega)\right)=2\beta F_{\rm static}(\omega),
$$
whose solution is
\begin{equation}
F_{\rm static}(\omega)=C\dfrac{e^{-4\beta \sqrt{\ln\dfrac{\omega}{v_F}\beta d}}}{\omega
\sqrt{\ln\dfrac{\omega}{v_F}\beta d}},
\end{equation}
$C$ being a constant. Of course, this result can also be found by direct integration in
Eq.~(\ref{dst}).

Substituting the expression for $ F_{\rm static}(\omega)$ into Eq.~(\ref{red}), we get
Eq.~(\ref{FF}) of the main text. We see, that the CDW form-factor is diverging near the
threshold, and the divergency in a CLL is very strong, almost $\sim1/\epsilon$, which
would correspond to the limit $g\to0$ in the short-ranged case. The divergency in a CLL
is, nevertheless, integrable (since the expression in Eq.~(\ref{FF}) is the exact
differential).

The LL spectral function satisfies the integral equation, similar to Eq.~(\ref{main}).
Using the above procedure, we solve it and get the Eq.~(\ref{spect}) of the main text.

\newpage

\end{document}